\documentclass[structabstract]{aa}  

\usepackage{graphicx}
\usepackage{longtable}
\usepackage{natbib}
\bibpunct{(}{)}{;}{a}{}{,}
\usepackage[varg]{txfonts}
\defcitealias{stonkute12}{Paper~I}
\begin{document}
   \title{Stellar substructures in the solar neighbourhood}

   \subtitle{II. Abundances of neutron-capture elements in the kinematic Group 3 of the Geneva-Copenhagen survey}

   \author{E. Stonkut\.{e}\inst{1},
          G. Tautvai\v{s}ien\.{e}\inst{1},
            B. Nordstr\"{o}m\inst{2},
          \and
          R. \v{Z}enovien\.{e}\inst{1}
          }
   \institute{Institute of Theoretical Physics and Astronomy, Vilnius University,
              A. Gostauto 12, LT-01108 Vilnius, Lithuania\\
              \email{[edita.stonkute;grazina.tautvaisiene;renata.zenoviene]@tfai.vu.lt}
         \and
        Niels Bohr Institute, Copenhagen University, Juliane Maries Vej 30, DK-2100, Copenhagen, Denmark\\
             \email{birgitta@nbi.ku.dk}
             }

   \date{Received February 00, 2013; accepted March 00, 2013}

 
  \abstract
   {The evolution of chemical elements in a galaxy is linked to its star formation history. Variations in star formation history 
   are imprinted in the relative abundances of chemical elements produced in different supernova events and asymptotic giant branch stars.}
   {We determine detailed elemental abundances of s- and r-process elements in stars belonging to Group~3 of the Geneva-Copenhagen 
   survey and compare their chemical composition 
with Galactic disc stars. The aim is to look for possible chemical signatures that might give information about the formation history 
of this kinematic group of stars, which is suggested to correspond to remnants of disrupted satellites.}
   {High-resolution spectra were obtained with the FIES spectrograph at the Nordic Optical Telescope, La Palma, and were analysed
with a differential model atmosphere method. Comparison stars were observed and analysed with the same method.}
   {Abundances of chemical elements produced mainly by the s-process are similar to those in the Galactic thin-disc dwarfs 
     of the same metallicity, while abundances of chemical elements produced predominantly by the r-process are overabundant. 
     The similar elemental abundances are observed in Galactic thick-disc stars. }
{The chemical composition together with the kinematic properties and ages of stars in  
Group~3 of the Geneva-Copenhagen survey support a gas-rich satellite merger scenario as the most 
likely explanation for the origin.
The similar chemical composition of stars in Group~3 and the thick-disc stars might suggest that their 
formation histories are linked.}

               
\titlerunning{Stellar substructures in the solar neighbourhood. II.}
\authorrunning{E. Stonkut\.{e} et al.} 
   \maketitle
\section{Introduction}

Our understanding of the global star formation history of the Milky Way galaxy remains incomplete. We need 
to determine the detailed age and spatial distributions, space motions, and elemental 
abundances both globally and for every substructure. It is important to 
understand effects of accreted satellites, how they depend on the time of accretion, 
their initial orbits, masses and density profiles, since these factors impose different 
scenarios of the tidal disruption of satellites and the distribution of debris to different 
Galactic components (cf. \citealt{wyse09}; \citealt{vanderkruit11}, and references therein).

A number of stellar streams, moving and kinematic groups were identified in our Galaxy 
(\citealt{zuckerman04, helmi08, klement09, sesar12}, and references therein). 
Some of them are suspected to originate from accreted satellites.
Signatures of past accretions in the Milky Way may be identified from correlations 
between stellar orbital parameters, such as apocentre (A), pericentre (P), and \textit{z}-angular momentum ($L_z$), 
the so-called APL space. \citet{helmi06} identified three new coherent groups of stars in the Geneva-Copenhagen 
survey \citep{nordstrom04} and suggested that those 
might correspond to remains of disrupted satellites. In the kinematic \textit{U--V} plane, the investigated stars are 
distributed in a banana-shape, whereas the disc stars define a centrally concentrated clump. At the same time, 
in the \textit{U--W} plane the investigated stars populate mostly the outskirts of the distributions. Both the \textit{U} 
and \textit{W} distributions are very symmetric. The investigated stars have a lower mean galactic rotational velocity than the Milky Way disc stars in the \textit{W--V} plane. These characteristics are typical for stars 
associated with accreted satellite galaxies \citep{helmi08, villalobos09}. 
Stars in the identified groups also cluster around regions of roughly constant eccentricity 
($0.3 \le \epsilon < 0.5$).

In \citeauthor{stonkute12} (2012, Paper~I), we started to investigate the detailed chemical composition of stars belonging 
to kinematic groups of the Geneva-Copenhagen survey \citep{helmi06}. Group~3, which we decided to investigate first, 
is the most metal-deficient and consists of 68 stars. It differs from the other two groups by slightly different kinematics, 
particularly in the vertical ($z$) direction. 

From high-resolution spectra, we measured abundances of iron group and $\alpha$-elements 
in 21 stars. All stars in Group~3 except one have a similar metallicity, their age is of about 12~Gyr. The average 
[Fe/H]\footnote{We use the customary spectroscopic notation
[X/Y]$\equiv \log_{10}(N_{\rm X}/N_{\rm Y})_{\rm star} -\log_{10}(N_{\rm X}/N_{\rm Y})_\odot$.} value of 
the 20 stars is $-0.69\pm 0.05$~dex. All stars are overabundant in oxygen and $\alpha$-elements compared with Galactic thin-disc 
dwarfs and the Galactic evolution model \citep{pagel95}. This abundance pattern has similar characteristics 
as the Galactic thick disc. The similar chemical composition of stars in Group~3 
and the thick-disc stars might suggest that their formation histories are linked.
The kinematic properties of this stellar group fit a gas-rich satellite merger scenario \citep{brook04, brook05, dierickx10, wilson11, dimatteo11}.
 
The neutron-capture elements as well as $\alpha$-elements are very sensitive indicators of galactic evolution 
(\citealt{pagel97, tautvaisiene07, sneden08, tolstoy09, ting12}, and references therein). If stars have been formed in different 
environments, they normally have different element-to-iron ratios for a given metallicity. 
In this work, we determine abundances of s- and r-process elements by means of high-resolution spectroscopy for stars in Group~3. 

The r-process, which requires a high neutron flux level (with many n-captures 
over a timescale of a fraction of a second), is believed to occur in supernova explosions. The s-process, which in contrast
requires a lower neutron flux (with a typical n-capture taking many years), is generally thought to occur during the
double-shell burning phase of low- ($1-3~M_\odot$) or intermediate-mass ($4-7~M_\odot$) thermally pulsing asymptotic giant branch
(AGB) stars.

We present n-capture element abundances 
(specifically Y, Zr, Ba, La, Ce, Pr, Nd, Sm, and Eu) for 21
stars of Group~3 and six comparison stars and compare them with the Galactic disc pattern. 

\section{Observations and analysis}

Echelle spectra of the programme and comparison stars were obtained with the high-resolution FIbre-fed Echelle Spectrograph (FIES) 
on the Nordic Optical 2.5 m telescope. This spectrograph gives spectra of resolving power $R\approx68\,000$ in the wavelength range 
of 3680--7270~{\AA}. The spectra were exposed to reach a signal-to-noise ratio higher than 100. Reductions of CCD images were made 
with the FIES pipeline FIEStool\footnote{http://www.not.iac.es/instruments/fies/fiestool}, which performs a complete reduction.

The spectra were analysed using a differential model atmosphere technique. The spectral synthesis was performed for the abundance determinations 
of all investigated chemical elements.

The program BSYN, developed at the Uppsala Astronomical Observatory, was used to carry out the calculations of synthetic spectra. 
A set of one-dimensional, hydrostatic, plane-parallel, line-blanketed, constant-flux LTE model atmospheres 
\citep{gustafsson08} was taken from the {\sc MARCS} stellar model atmosphere and flux 
library.\footnote{http://marcs.astro.uu.se}

Calibrations to the solar spectrum \citet{kurucz05} were made for all the spectral regions investigated. For this purpose we
used the solar model-atmosphere from the set calculated in Uppsala with a microturbulent velocity of 0.8~${\rm km~s}^{-1}$, as derived from
Fe\,{\sc i} lines. The atomic oscillator strengths for stronger lines of iron and
other elements were taken from \citet{gurtovenko89}.
The Vienna Atomic Line Database \citep[VALD;][]{piskunov95} was
extensively used in preparing the input data for the calculations.
In addition to thermal and microturbulent Doppler broadening of lines, atomic 
line broadening by radiation damping and van der Waals damping were considered 
in calculating the abundances. Radiation damping parameters of lines were taken from the VALD database. 
In most cases the hydrogen pressure damping of metal lines was treated using 
the modern quantum mechanical calculations by \citet{anstee95}, 
\citet{barklem97}, and \citet{barklem98}. 
When using the \citet{unsold55} approximation, correction factors to the classical 
van der Waals damping approximation by widths 
$(\Gamma_6)$ were taken from \citet{simmons82}. For all other species a correction factor 
of 2.5 was applied to the classical $\Gamma_6$ $(\Delta {\rm log}C_{6}=+1.0$), 
following \citet{mackle75}. For lines stronger than $W=100$~m{\AA} in the solar spectrum the correction factors 
were selected individually by inspecting the solar spectrum.

Effective temperature, gravity, [Fe/H], and microturbulent velocity values of the programme and comparison stars 
have been taken from \citetalias{stonkute12} where these values were derived using spectroscopic methods. 
A stellar rotation was taken into account using values of \textit{v}\,sin\,\textit{i} from \citet{holmberg07}.

 
   \begin{table}
   	\centering
         \caption{Adopted solar elemental abundances.} 
        
        \label{solar_abund}
       
         \begin{tabular}{lcclc}
            \hline
	    \hline
            \noalign{\smallskip}
Element & log(X/H)+12 & &Element & log(X/H)+12\\ 
            \noalign{\smallskip}
            \hline
            \noalign{\smallskip}
Fe      & 7.50  & &     Ce     & 1.58 \\
Y       & 2.24  & &	Pr     & 0.71 \\
Zr      & 2.60  & &     Nd     & 1.50 \\
Ba      & 2.13  & &	Sm     & 1.01 \\
La      & 1.17  & &	Eu     & 0.51 \\

            \hline
         \end{tabular}
     
\tablefoot{The solar abundances were taken from \cite{grevesse00}.}

   \end{table}

 
  \begin{figure}
   \centering
   \includegraphics[width=\hsize]{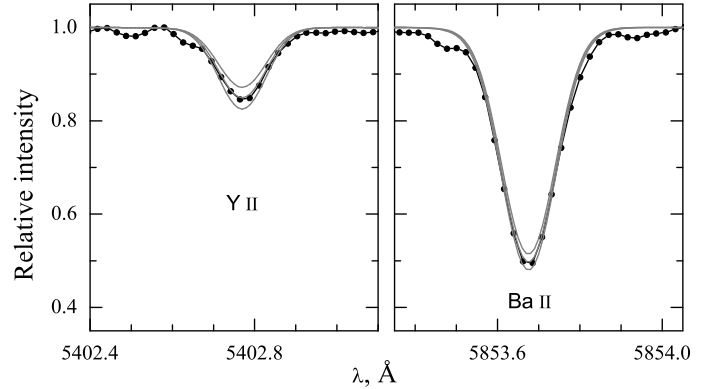}
      \caption{Synthetic spectrum fit to the yttrium line at 5402 $\AA$ and barium
 line at 5853 $\AA$. The observed spectrum for the programme star HD 204848 is shown as dots. The grey lines are
synthetic spectra with  [Y/Fe] = 0.33 $ \pm $ 0.10 and [Ba/Fe] = 0.25 $ \pm $ 0.10, respectively.}
         \label{YBa}
   \end{figure}

Abundances of the investigated chemical elements were determined from up to seven Y\,{\sc ii} lines at 4883.7, 4982.1, 
5087.4, 5200.4, 5289.8, 5402.8, and 5728.9 $\AA$; from  the Zr\,{\sc i} lines at 4687.8, 4772.3, 4815.6, 5385.1, 6134.6, 6140.5, 
and 6143.2 $\AA$ and Zr\,{\sc ii} lines at 5112.3 and 5350.1 $\AA$; from the Ba\,{\sc ii} line at 5853.7 $\AA$ with the hyperfine 
structure (HFS) and isotopic composition adopted from \citet{mcwilliam98}. 
The lanthanum abundance was determined from the La\,{\sc ii} lines at 4662.5, 4748.7, 5123.0, and 6390.5 $\AA$. 
To analyse the 4662.5, 5123.0, and 6390.5 $\AA$ 
lines, we adopted the log\,$gf$ from \citet{lawler01a} and HFS patterns from \citet{ivans06}.  The HFS patterns  
were not provided for the La\,{\sc ii} 
line at 4748.7 $\AA$. This line is very weak, so the broadening by hyperfine splitting can be neglected.
Up to five Ce\,{\sc ii} lines at 5274.2, 
5330.5, 5512.0, 5610.3, and 6043.4 $\AA$ were used to determine the abundance of cerium. 
The praseodymium abundance was determined from the Pr\,{\sc ii} lines at 5259.7 and 5322.8 $\AA$ with the information on     
HFS taken from \citet{sneden09}. 
We investigated Nd\,{\sc ii} lines at 4811.3, 5130.6, 5255.5, 5276.9, 5293.2, 5319.8, and 5385.9 $\AA$;   Sm\,{\sc ii} lines at 4467.3, 4577.7, and 
4791.6 $\AA$. For the Sm\,{\sc ii} line at 4467.3 $\AA$ the log\,$gf$ was taken from \citet{lawler06}, and the HFS patterns  from \citet{roederer08}. 
For the remaining two lines a hyperfine structure was not taken into account since these lines are 
very weak and their hyperfine splitting can be neglected \citep[cf.][]{mishenina13}.  
The abundance of europium was determined from the Eu\,{\sc ii} lines at 4129.7 and 6645.1 $\AA$. The log\,$gf$ values  for the Eu\,{\sc ii} lines and 
isotope fractions were 
adopted from \citet{lawler01b}.  An information on the HFS pattern for the Eu\,{\sc ii} line at 4129.7 $\AA$ 
was taken from \citet{ivans06}, and for the line at 6645.1 $\AA$  from \citet{biehl76}. A partial blending of the Eu\,{\sc ii} line 6645.1 $\AA$ with 
weak Si\,{\sc i} and Cr\,{\sc i} lines at 6645.21 $\AA$ was taken into account.


 \begin{figure}
   \centering
   \includegraphics[width=\hsize]{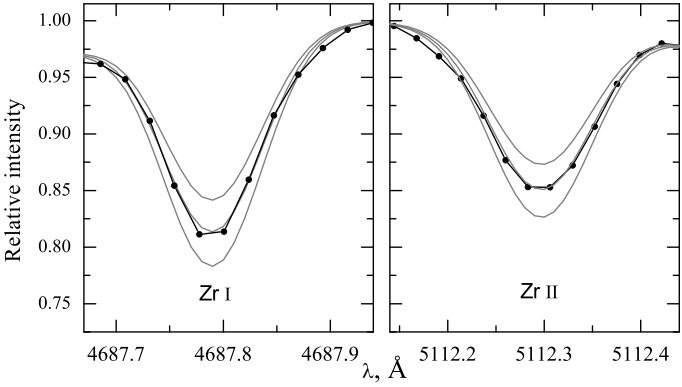}
      \caption{Synthetic spectrum fit to the Zr\,{\sc i} line at 4688 $\AA$ and Zr\,{\sc ii} line at 5112 $\AA$.
     The grey lines are
synthetic spectra with [Zr\,{\sc i}/Fe] = 0.40 $ \pm $ 0.10 and [Zr\,{\sc ii}/Fe] = 0.38 $ \pm $ 0.10, respectively. The observed spectrum for the programme 
star HD 204848 is shown as dots. }
         \label{ZrZr}
   \end{figure}


 \begin{figure}
   \centering
   \includegraphics[width=\hsize]{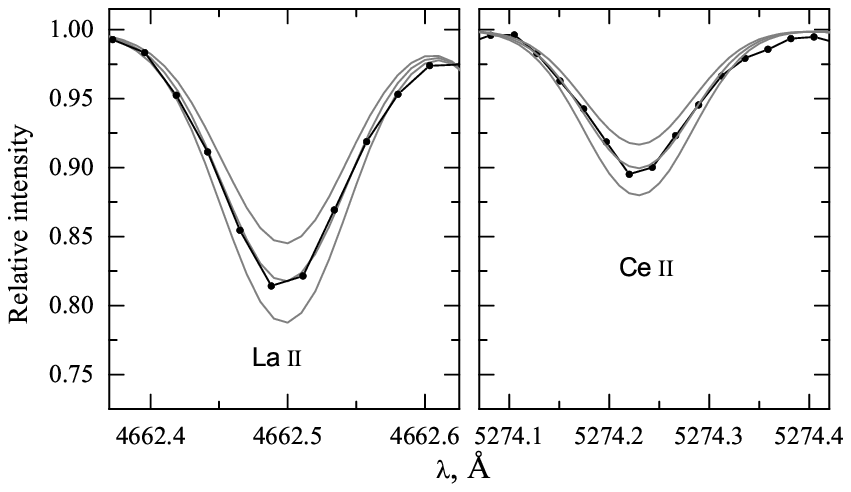}
      \caption{Synthetic spectrum fit to the lanthanum line at 4662 $\AA$ and cerium line at 5274 $\AA$. The observed spectrum for the programme star HD 204848 is 
      shown as dots. The grey lines are synthetic spectra with [La/Fe] = 0.20 $ \pm $ 0.10 and [Ce/Fe] = 0.10 $ \pm $ 0.10, respectively.}
         \label{LaCe}
   \end{figure}



 \begin{figure}
   \centering
   \includegraphics[width=\hsize]{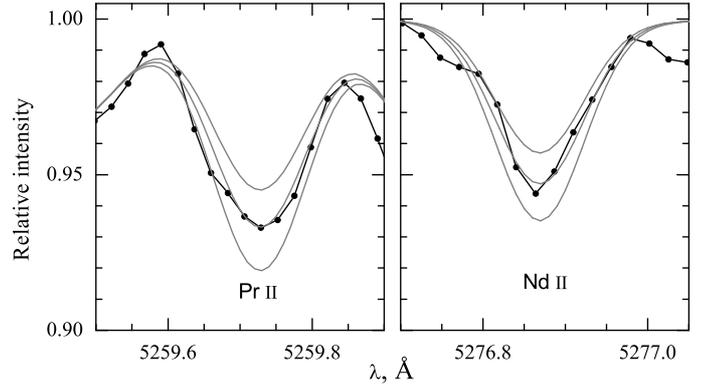}
      \caption{Synthetic spectrum fit to the praseodymium line at 5259 $\AA$  and neodymium line at 5276 $\AA$.
     The observed spectrum for the programme star HD 204848 is shown as dots. The grey lines are
synthetic spectra with [Pr/Fe] = 0.20 $ \pm $ 0.10 and [Nd/Fe] = 0.38 $ \pm $ 0.10, respectively.}
         \label{PrNd}
   \end{figure}

 \begin{figure}
   \centering
   \includegraphics[width=\hsize]{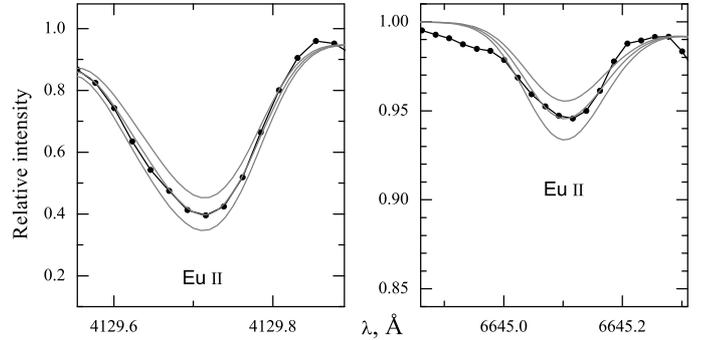}
      \caption{Synthetic spectrum fit to the Eu\,{\sc ii} lines at 4129 and 6645 $\AA$. The grey lines are
synthetic spectra with [Eu/Fe] = 0.33 $ \pm $ 0.10 and [Eu/Fe] = 0.37 $ \pm $ 0.10 for these two lines, respectively.
The observed spectrum for the programme star HD 170737 is shown as dots. }
         \label{EuEu}
   \end{figure}


The line parameters in the intervals of spectral syntheses were compiled from the VALD database. 
All log~$gf$ values were calibrated to fit to the solar spectrum by \cite{kurucz05} with solar abundances from \cite{grevesse00}. 
The adopted solar abundances for the investigated chemical elements are 
presented in Table \ref{solar_abund}.

Several fits of the synthetic line profiles to the observed
spectra are shown in Figs. \ref{YBa}, \ref{ZrZr}, \ref{LaCe}, \ref{PrNd}, and \ref{EuEu}. The best-fit abundances were determined by eye.

 \subsection{Estimation of uncertainties}

The sensitivity of oxygen, alpha, and Fe-peak element abundances to stellar atmospheric parameters were described in \cite{stonkute12}.
The sensitivity of the abundance estimates to changes in the atmospheric parameters by the assumed errors $\Delta$[El/H]
are illustrated for the star HD\,224930 (Table~\ref{Sens}).

The scatter of the deduced abundances from different spectral lines $\sigma$
gives an estimate of the uncertainty due to the random errors. The mean value 
of  $\sigma$ is 0.05~dex, thus the uncertainties in the derived abundances that 
are the result of random errors amount to approximately this value. 

\subsection{Comparison with other studies}


   \begin{table}
	\centering
      \caption{Effects on derived abundances resulting from model changes for the star HD\,224930.} 
         \label{Sens}
    
         \begin{tabular}{lrrcc}
            \hline
	    \hline
            \noalign{\smallskip}
Ion & ${ \Delta T_{\rm eff} }\atop{ +100 {\rm~K} }$ & 
            ${ \Delta \log g }\atop{ +0.3 }$ & 
            ${ \Delta v_{\rm t} }\atop{ +0.3 {\rm km~s}^{-1}}$ & Total\\ 
            \noalign{\smallskip}
            \hline
            \noalign{\smallskip}
Y\,{\sc ii}       & 	0.02	 & 	0.09	 & 	$-0.09$	 & 	0.13	  \\
Zr\,{\sc i}       & 	0.11	 & 	0.02	 & 	0.01	 & 	0.11	  \\
Zr\,{\sc ii}      & 	0.01	 & 	0.12	 & 	0.01	 & 	0.12	  \\
Ba\,{\sc ii}      & 	0.06	 & 	0.09	 & 	$-0.09$	 & 	0.14	  \\
La\,{\sc ii}      & 	0.04	 & 	0.11	 & 	0.01	 & 	0.12	  \\
Ce\,{\sc ii}      & 	0.03	 & 	0.10	 & 	0.01	 & 	0.10	  \\
Pr\,{\sc ii}      & 	0.04	 & 	0.09	 & 	0.01	 & 	0.10	  \\
Nd\,{\sc ii}      & 	0.04	 & 	0.11	 & 	0.00	 & 	0.12	  \\
Sm\,{\sc ii}      &     0.04     &      0.11     &     $-0.01$   &      0.11      \\
Eu\,{\sc ii}      & 	0.04	 & 	0.11	 & 	0.01	 & 	0.12	  \\

            \hline
         \end{tabular}
    \flushleft
\tablefoot{The table entries show the effects on the logarithmic abundances relative to hydrogen,
$\Delta$[El/H].}
   \end{table}

Some stars from our sample were previously investigated by other authors. In Table~\ref{table:3} we present a comparison with results by
\citet{nissen10} and \citet{reddy06}, who investigated several stars in common with our work.
Yttrium and barium for six thin-disc stars were analysed previously in
\citet{edvardsson93}. The comparison is presented in Table~\ref{table:3}. 
Our [El/Fe] values agree very well with 
other studies. The comparison for other chemical elements can be found in Paper~I.

\begin{table}
\begin{minipage}{80mm}
\caption{Comparison with previous studies.}
\label{table:3} 
\begin{tabular}{lrrrrrr}
\hline\hline   
         &\multicolumn{2}{c}{Our--Nissen} & \multicolumn{2}{c}{Our--Reddy} & \multicolumn{2}{c}{Our--Edvardsson}\\
Quantity & Diff.  & $\sigma$ &  Diff.    & $\sigma$  &  Diff.    & $\sigma$\\
\hline
${\rm [Fe/H]}$  &  0.03    & 0.04    & 0.06 	& 0.07 &  0.10    & 0.04 \\
${\rm[Y/Fe]}$   & 0.07     & 0.06     & 0.01 	& 0.09 & $-0.06$   & 0.11\\
${\rm[Ba/Fe]}$  & 0.13     & 0.05    & 0.11 	& 0.10  &$-0.04$    & 0.07 \\
${\rm[Ce/Fe]}$  & ...     & ...      & 0.01 	& 0.10  & ...     & ...\\
${\rm[Nd/Fe]}$  & ...     & ...      & $-0.05$ 	& 0.11  & ...     & ...\\
${\rm[Eu/Fe]}$  &...      & ...      & 0.08 	& 0.09  & ...     & ...\\
\hline
\end{tabular}
\end{minipage}
\tablefoot{Mean differences and standard deviations of the abundance ratios [El/Fe] for
four stars of Group~3 in common with \citet{nissen10}, for seven Group~3 stars in common with \citet{reddy06}, 
and for six thin-disc stars in common with \citet{edvardsson93}.}
\end{table}

\section{Results and discussion}

The elemental-to-iron abundance ratios derived for the programme and comparison stars are presented in Table \ref{table:results1}. 
We recall that BD +35 3659 is not a member of Group~3 and can be considered as a comparison star. 
For convenience we also present [Fe/H] values determined for these stars in \citetalias{stonkute12}. The number of lines used for the 
abundance determination and the line-to-line scatter ($\sigma$) are presented as well.

 \begin{figure}
\resizebox{\hsize}{!}{\includegraphics{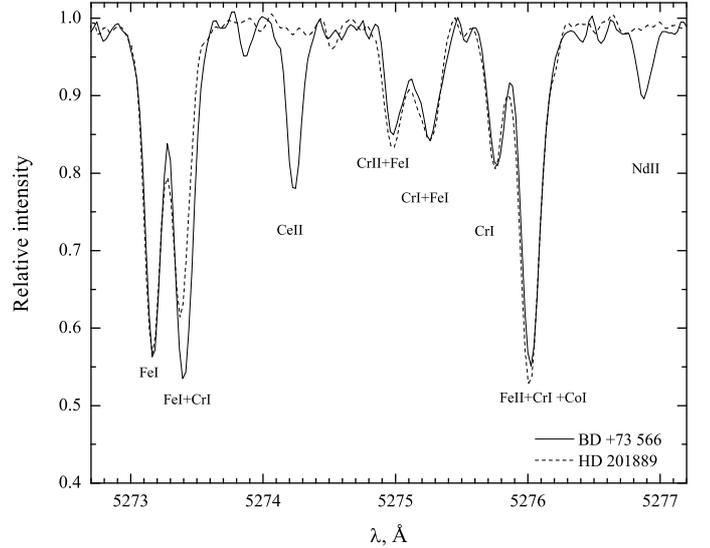}}
     \caption{The NOT-FIES spectra of the programme stars HD\,201889 and BD\,$+$73\,566. These two spectra  
     are over-plotted to see the difference in spectral lines of elements produced in s-and r-processes while lines of other chemical 
     elements are similar.}
       \label{s_rich}
   \end{figure}


 \begin{table*}
 \centering
\caption{Elemental abundances of Fe and neutron capture elements for the programme and comparison stars.}
\label{table:results1} 
\begin{tabular}{lcrccrccrccrrcccccccccc}
\hline\hline
\noalign{\smallskip}
Star & [Fe/H] & [Y/Fe] & $\sigma$ & n & [Zr\,{\sc i}/Fe] & $\sigma$ & n & [Zr\,{\sc ii}/Fe] & $\sigma$ & n & [Ba/Fe] & [La/Fe] & $\sigma$ & n \\

 \hline
\noalign{\smallskip}							
\object{HD 967}     	&	$	-0.62	$	&	$	0.06	$	&	$	0.05	$	&	4	&	$	0.24	$	&	$	0.03	$	&	4	&	$	0.24	$	&	$	0.00	$	&	2	&	$	-0.19	$	&	$	0.08	$	&	$	0.05	$	&	2	\\
\object{HD 17820}    	&	$	-0.57	$	&	$	-0.04	$	&	$	0.04	$	&	4	&	$	0.09	$	&	$	...	$	&	1	&	$	0.09	$	&	$	0.03	$	&	2	&	$	-0.10	$	&	$	0.07	$	&	$	0.03	$	&	3	\\
\object{HD 107582}   	&	$	-0.62	$	&	$	0.06	$	&	$	0.03	$	&	3	&	$	-0.05	$	&	$	0.07	$	&	2	&	$	-0.05	$	&	$	0.07	$	&	2	&	$	...	$	&	$	0.07	$	&	$	0.04	$	&	2	\\
\object{BD +73 566}  \tablefootmark{(a)}	&	$	-0.91	$	&	$	0.88	$	&	$	0.10	$	&	6	&	$	1.02	$	&	$	0.05	$	&	4	&	$	0.88	$	&	$	0.00	$	&	2	&	$	1.62	$	&	$	1.21	$	&	$	0.04	$	&	3	\\
\object{BD +19 2646} 	&	$	-0.68	$	&	$	-0.12	$	&	$	0.06	$	&	3	&	$	0.10	$	&	$	0.00	$	&	2	&	$	0.13	$	&	$	0.00	$	&	2	&	$	...	$	&	$	0.08	$	&	$	...	$	&	1	\\
\object{HD 114762}   	&	$	-0.67	$	&	$	-0.03	$	&	$	0.09	$	&	4	&	$	...	$	&	$	...	$	&	...	&	$	-0.05	$	&	$	...	$	&	1	&	$	-0.10	$	&	$	0.00	$	&	$	0.05	$	&	3	\\
\object{HD 117858}   	&	$	-0.55	$	&	$	-0.13	$	&	$	0.10	$	&	4	&	$	0.12	$	&	$	0.00	$	&	3	&	$	...	$	&	$	...	$	&	...	&	$	-0.11	$	&	$	-0.01	$	&	$	0.08	$	&	2	\\
\object{BD +13 2698} 	&	$	-0.74	$	&	$	-0.08	$	&	$	0.09	$	&	4	&	$	0.18	$	&	$	0.04	$	&	2	&	$	0.18	$	&	$	0.00	$	&	2	&	$	-0.20	$	&	$	0.10	$	&	$	0.06	$	&	2	\\
\object{BD +77 0521} 	&	$	-0.50	$	&	$	-0.12	$	&	$	0.00	$	&	3	&	$	...	$	&	$	...	$	&	...	&	$	...	$	&	$	...	$	&	...	&	$	...	$	&	$	...	$	&	$	...	$	&	...	\\
 \object{HD 126512}   	&	$	-0.55	$	&	$	-0.13	$	&	$	0.10	$	&	4	&	$	0.07	$	&	$	0.00	$	&	2	&	$	0.07	$	&	$	0.00	$	&	2	&	$	-0.18	$	&	$	-0.06	$	&	$	0.01	$	&	3	\\
 \object{HD 131597}   	&	$	-0.64	$	&	$	0.06	$	&	$	0.07	$	&	5	&	$	0.07	$	&	$	0.05	$	&	4	&	$	0.09	$	&	$	...	$	&	1	&	$	-0.10	$	&	$	0.02	$	&	$	0.09	$	&	4	\\
 \object{BD +67 925}  	&	$	-0.55	$	&	$	-0.18	$	&	$	0.00	$	&	2	&	$	0.00	$	&	$	...	$	&	1	&	$	...	$	&	$	...	$	&	1	&	$	-0.07	$	&	$	...	$	&	$	...	$	&	...	\\
\object{HD 159482}   	&	$	-0.71	$	&	$	-0.11	$	&	$	0.03	$	&	4	&	$	...	$	&	$	...	$	&	...	&	$	...	$	&	$	...	$	&	...	&	$	-0.17	$	&	$	0.05	$	&	$	0.00	$	&	3	\\
\object{HD 170737}   	&	$	-0.68	$	&	$	-0.12	$	&	$	0.06	$	&	7	&	$	-0.03	$	&	$	0.04	$	&	7	&	$	-0.03	$	&	$	...	$	&	1	&	$	...	$	&	$	0.05	$	&	$	0.01	$	&	3	\\
\object{BD +35 3659} \tablefootmark{(b)}	&	$	-1.45	$	&	$	-0.12	$	&	$	0.12	$	&	3	&	$	0.20	$	&	$	...	$	&	1	&	$	0.20	$	&	$	...	$	&	1	&	$	-0.12	$	&	$	0.20	$	&	$	...	$	&	1	\\
\object{HD 201889}   	&	$	-0.73	$	&	$	0.05	$	&	$	0.05	$	&	6	&	$	0.21	$	&	$	0.08	$	&	3	&	$	0.21	$	&	$	...	$	&	1	&	$	-0.06	$	&	$	0.04	$	&	$	0.07	$	&	4	\\
\object{HD 204521}   	&	$	-0.72	$	&	$	0.01	$	&	$	0.05	$	&	6	&	$	0.13	$	&	$	0.08	$	&	3	&	$	0.13	$	&	$	0.00	$	&	2	&	$	-0.10	$	&	$	0.09	$	&	$	0.01	$	&	3	\\
\object{HD 204848}   	&	$	-1.03	$	&	$	0.28	$	&	$	0.08	$	&	7	&	$	0.38	$	&	$	0.06	$	&	7	&	$	0.42	$	&	$	0.05	$	&	2	&	$	0.25	$	&	$	0.25	$	&	$	0.05	$	&	4	\\
\object{HD 212029}   	&	$	-0.98	$	&	$	0.18	$	&	$	0.05	$	&	4	&	$	...	$	&	$	...	$	&	...	&	$	...	$	&	$	...	$	&	...	&	$	0.24	$	&	$	0.22	$	&	$	0.00	$	&	3	\\
\object{HD 222794}   	&	$	-0.61	$	&	$	-0.05	$	&	$	0.05	$	&	5	&	$	0.12	$	&	$	0.06	$	&	3	&	$	...	$	&	$	...	$	&	...	&	$	-0.10	$	&	$	0.04	$	&	$	0.02	$	&	3	\\
\object{HD 224930}   	&	$	-0.71	$	&	$	0.05	$	&	$	0.05	$	&	5	&	$	0.13	$	&	$	0.04	$	&	2	&	$	0.13	$	&	$	...	$	&	1	&	$	-0.17	$	&	$	0.07	$	&	$	0.03	$	&	3	\\

\noalign{\smallskip}
\hline
 \noalign{\smallskip}	
\object{HD 17548}    	&	$	-0.49	$	&	$	-0.12	$	&	$	0.06	$	&	5	&	$	0.00	$	&	$	0.00	$	&	2	&	$	0.00	$	&	$	0.00	$	&	1	&	$	-0.10	$	&	$	-0.07	$	&	$	0.05	$	&	3	\\
\object{HD 150177}   	&	$	-0.50	$	&	$	0.02	$	&	$	0.07	$	&	5	&	$	0.10	$	&	$	0.09	$	&	3	&	$	0.06	$	&	$	0.01	$	&	2	&	$	-0.09	$	&	$	0.03	$	&	$	0.06	$	&	3	\\
\object{HD 159307}   	&	$	-0.60	$	&	$	0.05	$	&	$	0.07	$	&	2	&	$	...	$	&	$	...	$	&	...	&	$	0.06	$	&	$	...	$	&	1	&	$	-0.08	$	&	$	...	$	&	$	...	$	&	...	\\
\object{HD 165908}   	&	$	-0.52	$	&	$	-0.12	$	&	$	0.06	$	&	5	&	$	-0.05	$	&	$	0.00	$	&	2	&	$	-0.05	$	&	$	...	$	&	1	&	$	-0.09	$	&	$	0.01	$	&	$	0.06	$	&	4	\\
\object{HD 174912}   	&	$	-0.42	$	&	$	-0.12	$	&	$	0.02	$	&	4	&	$	0.14	$	&	$	0.04	$	&	3	&	$	0.04	$	&	$	...	$	&	1	&	$	-0.07	$	&	$	-0.05	$	&	$	0.03	$	&	3	\\
\object{HD 207978}   	&	$	-0.50	$	&	$	-0.04	$	&	$	0.06	$	&	4	&	$	0.20	$	&	$	0.00	$	&	6	&	$	0.11	$	&	$	...	$	&	1	&	$	-0.10	$	&	$	-0.07	$	&	$	0.00	$	&	2	\\
            \noalign{\smallskip}
\end{tabular}
\begin{tabular}{lrccrccrccrccrccccccccc}
\hline\hline
\noalign{\smallskip}
Star & [Ce/Fe] & $\sigma$ & n & [Pr/Fe]  & $\sigma$ & n & [Nd/Fe]  & $\sigma$ & n & [Sm/Fe]  & $\sigma$ & n & [Eu/Fe] & $\sigma$ & n \\
 \hline
\noalign{\smallskip}		 
HD 967      	&	$	-0.02	$	&	$	0.04	$	&	2	&	$	0.36	$	&	$	0.08	$	&	2	&	$	0.13	$	&	$	0.05	$	&	4	&	$	0.32	$	&	$	0.03	$	&	3	&	$	0.39	$	&	$	0.06	$	&	2	\\
 HD 17820    	&	$	0.11	$	&	$	...	$	&	1	&	$	0.37	$	&	$	0.00	$	&	2	&	$	0.14	$	&	$	0.05	$	&	5	&	$	0.38	$	&	$	0.06	$	&	3	&	$	0.38	$	&	$	0.07	$	&	2	\\
 HD 107582   	&	$	-0.04	$	&	$	0.06	$	&	5	&	$	0.34	$	&	$	...	$	&	1	&	$	0.10	$	&	$	0.05	$	&	5	&	$	0.28	$	&	$	0.05	$	&	3	&	$	0.38	$	&	$	0.07	$	&	2	\\
 BD +73 566  	&	$	1.25	$	&	$	0.00	$	&	3	&	$	1.12	$	&	$	0.04	$	&	2	&	$	1.33	$	&	$	0.07	$	&	7	&	$	1.00	$	&	$	0.15	$	&	3	&	$	0.64	$	&	$	0.29	$	&	2	\\
 BD +19 2646 	&	$	0.07	$	&	$	0.05	$	&	3	&	$	0.40	$	&	$	...	$	&	1	&	$	0.11	$	&	$	0.04	$	&	5	&	$	0.22	$	&	$	0.03	$	&	3	&	$	0.30	$	&	$	...	$	&	1	\\
 HD 114762   	&	$	0.07	$	&	$	0.02	$	&	3	&	$	0.21	$	&	$	0.01	$	&	2	&	$	0.07	$	&	$	0.03	$	&	4	&	$	0.23	$	&	$	0.05	$	&	3	&	$	0.17	$	&	$	0.11	$	&	2	\\
 HD 117858   	&	$	0.15	$	&	$	0.00	$	&	3	&	$	0.32	$	&	$	0.03	$	&	2	&	$	-0.04	$	&	$	0.06	$	&	4	&	$	0.28	$	&	$	0.06	$	&	3	&	$	0.28	$	&	$	0.01	$	&	2	\\
 BD +13 2698 	&	$	-0.05	$	&	$	0.00	$	&	2	&	$	0.35	$	&	$	...	$	&	1	&	$	0.05	$	&	$	0.06	$	&	4	&	$	0.33	$	&	$	0.07	$	&	3	&	$	0.31	$	&	$	0.03	$	&	2	\\
 BD +77 0521 	&	$	...	$	&	$	...	$	&	...	&	$	...	$	&	$	...	$	&	...	&	$	...	$	&	$	...	$	&	...	&	$	...	$	&	$	...	$	&	...	&	$	...	$	&	$	...	$	&	...	\\
 HD 126512   	&	$	0.05	$	&	$	0.00	$	&	3	&	$	0.30	$	&	$	0.00	$	&	1	&	$	0.03	$	&	$	0.06	$	&	5	&	$	0.21	$	&	$	0.14	$	&	3	&	$	0.28	$	&	$	0.02	$	&	2	\\
 HD 131597   	&	$	-0.01	$	&	$	0.06	$	&	5	&	$	0.26	$	&	$	0.02	$	&	2	&	$	0.13	$	&	$	0.06	$	&	7	&	$	0.26	$	&	$	0.06	$	&	2	&	$	0.32	$	&	$	0.03	$	&	2	\\
 BD +67 925  	&	$	...	$	&	$	...	$	&	...	&	$	0.32	$	&	$	0.04	$	&	2	&	$	0.08	$	&	$	0.05	$	&	6	&	$	...	$	&	$	...	$	&	...	&	$	0.29	$	&	$	0.07	$	&	2	\\
 HD 159482   	&	$	0.05	$	&	$	0.00	$	&	2	&	$	0.40	$	&	$	...	$	&	1	&	$	0.12	$	&	$	0.06	$	&	5	&	$	0.30	$	&	$	0.06	$	&	3	&	$	0.33	$	&	$	0.11	$	&	2	\\
 HD 170737   	&	$	0.00	$	&	$	0.08	$	&	4	&	$	0.12	$	&	$	0.03	$	&	2	&	$	0.14	$	&	$	0.05	$	&	7	&	$	0.25	$	&	$	0.11	$	&	2	&	$	0.35	$	&	$	0.03	$	&	2	\\
 BD +35 3659 	&	$	...	$	&	$	...	$	&	...	&	$	...	$	&	$	...	$	&	...	&	$	0.25	$	&	$	0.00	$	&	3	&	$	0.37	$	&	$	...	$	&	1	&	$	0.55	$	&	$	...	$	&	1	\\
 HD 201889   	&	$	0.06	$	&	$	0.08	$	&	4	&	$	0.33	$	&	$	0.05	$	&	2	&	$	0.06	$	&	$	0.06	$	&	6	&	$	0.29	$	&	$	0.02	$	&	3	&	$	0.26	$	&	$	0.09	$	&	2	\\
 HD 204521   	&	$	0.06	$	&	$	0.09	$	&	2	&	$	0.39	$	&	$	0.00	$	&	2	&	$	0.10	$	&	$	0.05	$	&	5	&	$	0.30	$	&	$	0.05	$	&	3	&	$	0.40	$	&	$	0.04	$	&	2	\\
 HD 204848   	&	$	0.13	$	&	$	0.13	$	&	5	&	$	0.24	$	&	$	0.06	$	&	2	&	$	0.30	$	&	$	0.07	$	&	6	&	$	0.28	$	&	$	0.06	$	&	3	&	$	0.27	$	&	$	0.01	$	&	2	\\
 HD 212029   	&	$	...	$	&	$	...	$	&	...	&	$	...	$	&	$	...	$	&	...	&	$	0.25	$	&	$	0.09	$	&	3	&	$	0.50	$	&	$	0.00	$	&	2	&	$	0.52	$	&	$	0.10	$	&	2	\\
 HD 222794   	&	$	0.08	$	&	$	0.05	$	&	3	&	$	0.40	$	&	$	0.02	$	&	2	&	$	0.09	$	&	$	0.04	$	&	5	&	$	0.24	$	&	$	0.06	$	&	2	&	$	0.37	$	&	$	0.04	$	&	2	\\
 HD 224930   	&	$	0.08	$	&	$	0.11	$	&	2	&	$	0.40	$	&	$	...	$	&	1	&	$	0.01	$	&	$	0.02	$	&	3	&	$	0.31	$	&	$	0.06	$	&	3	&	$	0.30	$	&	$	0.08	$	&	2	\\
\noalign{\smallskip}
\hline
 \noalign{\smallskip}
HD 17548    	&	$	0.08	$	&	$	0.03	$	&	3	&	$	0.21	$	&	$	...	$	&	1	&	$	0.01	$	&	$	0.05	$	&	4	&	$	0.17	$	&	$	0.05	$	&	3	&	$	0.18	$	&	$	0.06	$	&	2	\\
 HD 150177   	&	$	0.05	$	&	$	0.00	$	&	2	&	$	0.15	$	&	$	...	$	&	1	&	$	0.07	$	&	$	0.06	$	&	6	&	$	0.15	$	&	$	0.02	$	&	3	&	$	0.20	$	&	$	...	$	&	1	\\
 HD 159307   	&	$	0.05	$	&	$	...	$	&	1	&	$	...	$	&	$	...	$	&	...	&	$	0.02	$	&	$	0.00	$	&	2	&	$	0.26	$	&	$	0.04	$	&	2	&	$	...	$	&	$	...	$	&	...	\\
 HD 165908   	&	$	0.02	$	&	$	0.06	$	&	3	&	$	0.21	$	&	$	0.01	$	&	2	&	$	0.02	$	&	$	0.09	$	&	6	&	$	0.12	$	&	$	0.03	$	&	3	&	$	0.13	$	&	$	0.07	$	&	2	\\
 HD 174912   	&	$	0.07	$	&	$	0.03	$	&	3	&	$	0.09	$	&	$	0.07	$	&	2	&	$	0.02	$	&	$	0.07	$	&	5	&	$	0.17	$	&	$	0.08	$	&	3	&	$	0.09	$	&	$	0.11	$	&	2	\\
 HD 207978   	&	$	0.06	$	&	$	...	$	&	1	&	$	...	$	&	$	...	$	&	...	&	$	0.10	$	&	$	0.00	$	&	3	&	$	0.17	$	&	$	0.03	$	&	3	&	$	0.16	$	&	$	0.04	$	&	2	\\

   \hline
            \noalign{\smallskip}

\end{tabular}
\tablefoot{
[Fe/H] are taken from \citetalias{stonkute12}. \tablefoottext{a}{The star is rich in chemical elements produced by s-and r-processes.} \tablefoottext{b}{Probably not a member of Group~3.}}
\end{table*}						


 \begin{figure}[h]
\resizebox{\hsize}{!}{\includegraphics{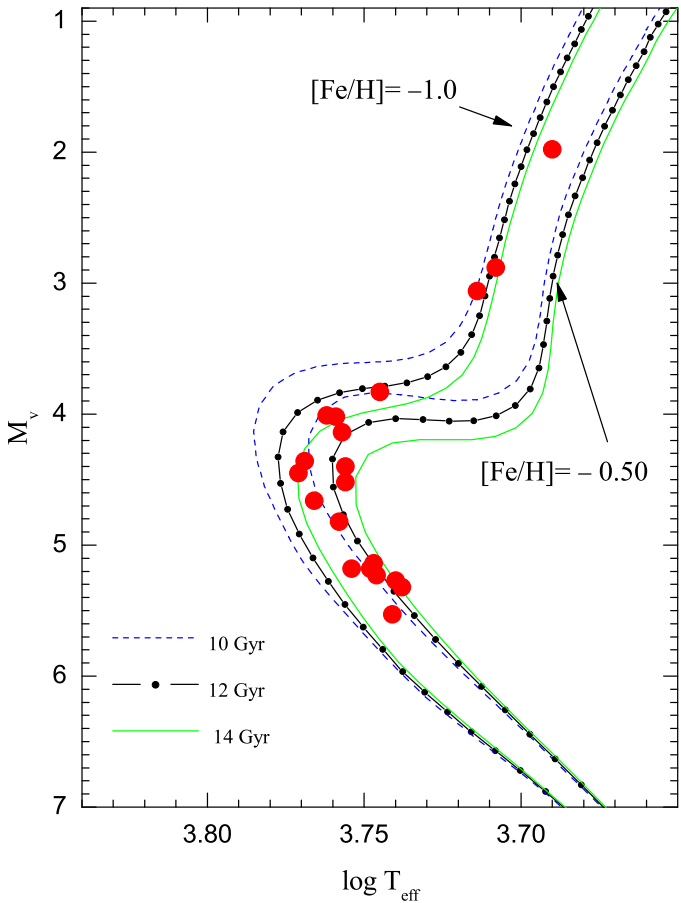}}
     \caption{HR diagram of the Group 3 stars. \citet{yi01} isochrones representing three different ages and metallicities are shown. The filled circles 
     correspond to the investigated stars with the spectroscopic effective temperatures taken from \citetalias{stonkute12}. Isochrones are with [$\alpha$/Fe ]= 0.4 dex. }
       \label{age}
   \end{figure}

The age distribution of the Group 3 stars is seen on a Hertzsprung-Russell (HR) diagram (Fig.~\ref{age}).  
The initial age evaluation for Group~3 was 14~Gyr \citep{holmberg07}. 
In Fig.~\ref{age} we plot the investigated stars 
of Group 3 with our spectroscopic effective temperatures \citepalias{stonkute12} and absolute magnitudes $M_{v}$ taken from \citet{holmberg09}. 
We used the Yonsei-Yale single 
stellar population library by \citet{yi01}, and its updated version by \citet{demarque04}. 
We plotted isochrones of three ages (10, 12, and 14 Gyr) and two metallicities [Fe/H] ($-1.0$ and $-0.50$). 
These two metallicities 
represent the minimum and maximum values of Group~3 stars. The mean metallicity of investigated Group~3 stars is equal to 
$-0.69\pm 0.05$~dex. The overall features of the HR diagram seem to be well reproduced by the 12~Gyr isochrone. We note that 
ages of stars belonging to the Arcturus and AF06 stellar streams also range from 10 to 14~Gyr \citep{ramya12}.

One star in Group~3 is rich in elements produced in s-and r-processes. As is seen from Table \ref{table:results1} and Fig.\,\ref{s_rich}, star BD\,$+$73\,566 has much stronger lines of elements produced in s-and r-processes and consequently much 
higher abundances of such elements. According to the definition of \citet{beers05}, BD\,$+$73\,566, with its ${\rm [Ba/Fe]}=1.62$ and ${\rm [Ba/Eu]}=0.98$,  
falls in the category of the s-process-enhanced stars.

\begin{figure*}
   \centering
   \includegraphics[width=0.85\textwidth]{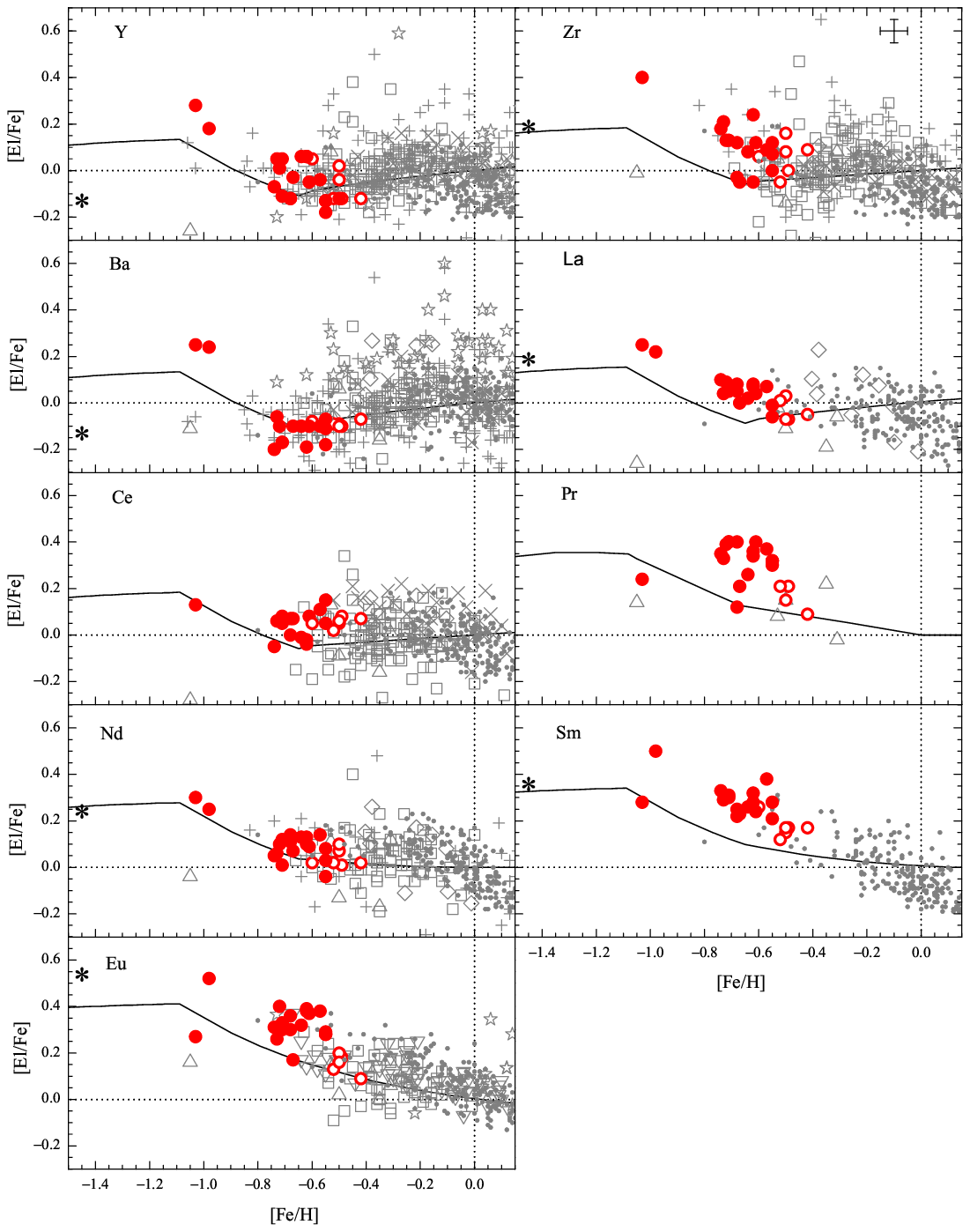}
     \caption{Elemental abundance ratios of stars in Group~3 (filled circles) and comparison stars (open circles). The star BD +35 3659, which 
     was removed from Group~3, is marked by an asterisk. For this comparison the data for Milky Way thin-disc dwarfs are plotted: 
     \citet[dots]{mishenina13}, \citet[crosses]{mashonkina07}, \citet[squares]{reddy06, reddy03}, \citet[diamonds]{brewer06}, \citet[stars]{bensby05}, 
     \citet[upside down triangles]{koch02}, \citet[triangles]{gratton94}, \citet[plus signs]{edvardsson93}. Solid lines show the 
     Galactic thin-disc chemical evolution models by \citet{pagel97}. Average uncertainties are shown in the box for zirconium.}
       \label{results}
   \end{figure*}

[El/Fe] ratios for the programme and comparison stars (except  BD\,$+$73\,566) are plotted in Fig.~\ref{results}. For the comparison 
we used six stars of the thin-disc observed in our work as well as data from other studies of thin-disc stars 
\citep{mishenina13, mashonkina07, reddy06, reddy03, brewer06, bensby05, koch02, gratton94, edvardsson93}. The comparison was made 
also with the Galactic thin disc chemical evolution models by \citet{pagel97}. 

In Fig.~\ref{results} we see that the abundances of yttrium and barium, which are produced mainly in the s-process, are the same as in the thin-disc stars. 
The abundances of chemical elements for which the r-process contribution is higher or dominating are higher than in the thin disc. This is the case for europium, samarium, and praseodymium. The element-to-iron ratios for these elements in Group~3 stars are higher than in the investigated comparison stars and other thin-disc 
stars. For zirconium, lanthanum, cerium, and neodymium, Group~3 and the comparison stars have approximately similar element-to-iron ratios. 

The similar pattern of n-capture element-to-iron ratios is observed in thick-disc stars 
\citep[e.g.][]{mashonkina00, prochaska00, tautvaisiene01, bensby05, reddy06, reddy08, mishenina13}.
A thick-disc-like overabundance of $\alpha$-elements was found in these stars as well 
\citepalias{stonkute12}. 
In Table~\ref{table:thick-disc} we show the comparison of mean [El/Fe] values for stars of Group~3 and thick-disc stars at the same metallicity interval $-0.8 < {\rm [Fe/H]} < -0.5$. In this metallicity interval 
lie almost all stars of Group~3. The comparison was made with stars from three thick-disc studies: ten stars lying in the mentioned metallicity interval were taken from 
\citet{bensby05}, 44 stars from \citet{reddy06}, and six stars from \citet{mishenina13}. Clearly, Group~3 stars and thick-disc stars are of similar chemical composition. The deviations do not exceed the uncertainties of chemical composition determinations.

As we outlined in \citetalias{stonkute12}, the similar chemical composition of stars in Group~3 and thick-disc stars 
might suggest that their formation histories are linked. 
In this context it is important to note that the recent chemical composition study of two stellar streams in 
the Galaxy also have shown thick-disc like abundances of elements. This was found in \citet{ramya12},
who investigated abundances of 20 elements in Arcturus and AF06 stellar streams. 
The authors referred to dynamical interactions 
within the Galaxy as the most probable scenario for the origin 
of these two streams. However, it is not clear 
how perturbations can create streams with such a very high velocity drag and at the same time exhibit very tight abundance trends.
\begin{table}
\begin{minipage}{80mm}
\caption{Comparison with thick-disc studies.}
\label{table:thick-disc} 
\begin{tabular}{lrrr}
\hline\hline   
[El/Fe]         & Group3 -- B05 & Group3 -- R06 & Group3 -- M13\\
\hline
${\rm [O/Fe]}$   	&	$	0.00	$	&	$	 ... 	$	&	$	-0.03	$	\\
${\rm [Mg/Fe]}$  	&	$	-0.01	$	&	$	0.02	$	&	$	0.01	$	\\
${\rm [Si/Fe]}$  	&	$	0.03	$	&	$	0.03	$	&	$	-0.03	$	\\
${\rm [Ca/Fe]}$  	&	$	0.06	$	&	$	0.10	$	&	$	0.03	$	\\
${\rm [Ti/Fe]}$  	&	$	0.05	$	&	$	0.09	$	&	$	 ...       	$	\\
${\rm[Y/Fe]}$    	&	$	0.05	$	&	$	-0.08	$	&	$	-0.08	$	\\
${\rm [Zr/Fe]}$  	&	$	 ...      	$	&	$	 ... 	$	&	$	-0.06	$	\\
${\rm[Ba/Fe]}$  	&	$	-0.10	$	&	$	0.00	$	&	$	-0.12	$	\\
${\rm [La/Fe]}$ 	&	$	 ...       	$	&	$	 ... 	$	&	$	0.04	$	\\
${\rm[Ce/Fe]}$  	&	$	 ...       	$	&	$	-0.02	$	&	$	0.04	$	\\
${\rm[Nd/Fe]}$  	&	$	 ...       	$	&	$	-0.11	$	&	$	-0.10	$	\\
${\rm [Sm/Fe]}$ 	&	$	 ...       	$	&	$	 ... 	$	&	$	0.06	$	\\
${\rm[Eu/Fe]}$  	&	$	-0.05	$	&	$	-0.03	$	&	$	-0.06	$	\\

\hline
\end{tabular}
\end{minipage}
\tablefoot{Differences of mean [El/Fe] values for
stars of Group~3 and thick-disc stars at the same metallicity interval $-0.8 < {\rm [Fe/H]} < -0.5$. 
B05 -- 10 stars from \citet{bensby05}; R06 -- 44 stars from \citet{reddy06}; M13 -- 6 stars from \citet{mishenina13}.}
\end{table}

Thus, as was pointed out by \citet{helmi06} and by us in \citetalias{stonkute12},
Group~3 can not be uniquely associated to a single traditional Galactic component.
The chemical composition together with the kinematic properties and ages of stars in the 
investigated Group~3 of the Geneva-Copenhagen survey support a gas-rich satellite merger scenario as most 
suitable for Group~3 origin.  A gas-rich satellite merger may be responsible 
for the formation of the Galactic thick-disc as well \citep{samland03, brook04, brook05, dierickx10, wilson11, dimatteo11}.


\section{Conclusions}

Using high-resolution spectra, we measured abundances of chemical elements produced by s- and r-processes 
in stars attributed to Group~3 of the Geneva-Copenhagen survey. This kinematically identified group of stars 
was suspected to be a remnant of a disrupted satellite galaxy.
Our main goal was to investigate the homogeneity of the chemical composition of the stars within the group and to 
compare them with Galactic disc stars. 

Our study of 20 stars in Group~3 shows the following:
  \begin{enumerate}

     \item The abundances of chemical elements produced mainly by the s-process are similar to those in the Galactic thin-disc dwarfs 
     of the same metallicity. 
     
     \item The abundances of chemical elements produced predominantly by the r-process are overabundant 
     in comparison with Galactic thin-disc dwarfs of the same metallicity. The most prominent overabundances are seen for europium, 
samarium, and praseodymium. 
  
    \item The chemical composition of stars in Group~3 is similar to the thick-disc stars, which might suggest that their 
formation histories are linked.
   
     \item BD +73 566 is an s-process-enhanced star.

     \item Group~3 consists of a 12-Gyr-old population.  
 
     \item The chemical composition together with the kinematic properties and ages of stars in the 
investigated Group~3 of the Geneva-Copenhagen survey support a gas-rich satellite merger scenario as the most 
suitable origin for Group~3. 

  \end{enumerate}

Other kinematic groups of the Geneva-Copenhagen survey will be analysed in forthcoming papers of this series.

\begin{acknowledgements}
The data are based on observations made with the Nordic Optical Telescope, operated on the island of La Palma jointly by Denmark, Finland, Iceland,
Norway, and Sweden, in the Spanish Observatorio del Roque de los Muchachos of the Instituto de Astrofisica de Canarias.  
The research leading to these results has received funding from the European Community's Seventh Framework Programme (FP7/2007-2013) under
grant agreement number RG226604 (OPTICON). BN acknowledges support from the Danish Research council. 
This research has made use of Simbad, VALD and NASA ADS databases. 
We thank the anonymous referee for insightful questions and comments.

\end{acknowledgements}

\end{document}